\documentstyle[prb,aps,twocolumn]{revtex}

\begin{document}
\draft
\title{A Quantum Chemical Approach to Cohesive Properties of NiO}
\author{
  Klaus Doll, Michael Dolg, Peter Fulde}
\address{
     Max-Planck-Institut f\"ur Physik komplexer Systeme,
          Bayreuther Str.~40, D-01187 Dresden, Germany }
\author{Hermann Stoll}

\address{
Institut f\"ur Theoretische Chemie, Universit\"at Stuttgart, D-70550 Stuttgart,
Germany
}

\maketitle

\begin{abstract}
We apply {\em ab-initio} quantum chemical methods to
calculate correlation effects on
cohesive properties of NiO, thereby extending a recently
proposed scheme to transition metal oxides with  partially filled $d$-bands.
We obtain good agreement with experiment for the cohesive energy and show that 
the deviation of the lattice constant at the Hartree-Fock level is
mainly due to van der Waals-like interactions. Correlations enhance the
stability of the 
magnetic ground state found at the Hartree-Fock level.
\end{abstract}

\pacs{ }

\narrowtext
\section{Introduction}
NiO has a long tradition in solid-state physics as a prototype of
an insulator with partially filled $d$-bands \cite{DeBoerVerwey}. The initial
explanation for its unusual behavior was given by Mott \cite{Mott} who 
argued that the Coulomb repulsion of the electrons in 
singly occupied nickel orbitals 
should completely suppress excitations of the $d \rightarrow d$ type.
This picture led to the notion of Mott insulators and was later refined
by Zaanen et al \cite{Zaanen} who characterized NiO as
a charge-transfer insulator with a valence band of 
oxygen $p$-type and a resulting  $p \rightarrow d$ gap.
The band structure and especially the gap have been in the center of  interest
and a considerable amount of work has been devoted to their determination 
\cite{Mattheiss,Terakura,Fujimori,Janssen,Svane,Shen,Anisimov,Manghi,Aryasetiawan,Igarashi,Towler}
(for reviews on NiO see, e.g., Refs. 
\onlinecite{Brandow,Huefner} and, more general, Ref.  \onlinecite
{Fuldebuch}).
Surprisingly, only little work has been done on cohesive properties and
we are aware of only one result from density functional theory (DFT) 
\cite{Yamashita}.
At the Hartree-Fock (HF) level, a very careful and
extensive investigation has been performed \cite{Towler}. 
In the present paper,
we want to address the question how electron 
correlations affect ground-state
properties. 

In the field of atoms and molecules, the most accurate methods are those
of quantum chemistry whereas  in solid-state physics the dominant method
is DFT.
However, it has been demonstrated by using local operators 
combined with an incremental method that high-quality
results can be obtained from a quantum-chemical approach to
semiconductors
\cite{StollBeate}  and ionic compounds \cite{DDFS}.
Transition metals
are still a challenge in quantum chemistry and even for the NiO molecule
extensive calculations are necessary to obtain
agreement with  experiment.
We want to show, however,  that 
an accurate determination of correlation effects is possible nowadays, using
the afore-mentioned incremental scheme, even
for solids containing transition-metal compounds. 
In addition, we want to compare some relevant atomic
and molecular  results from {\em ab-initio}
 quantum chemistry
to those of DFT. For this purpose, we performed test calculations with
the local density approximation (LDA) as well as with the gradient-corrected
functional of Becke \cite{Becke},
Lee, Yang and Parr (BLYP) \cite{LYP}. 

\section{The Method}
\subsection{Incremental scheme}
The incremental scheme uses the fact that electron correlations are a
{\em local} property.
The 
first step is to calculate the correlation energy 
$\epsilon(A)$
of groups of localized
orbitals (the so-called one-body increments) \cite{Graphit}. 
The 
correlation energy of two groups of orbitals $\epsilon(AB)$ can then
be expressed as 
\begin{eqnarray*}
\epsilon(AB)=\epsilon(A)+\epsilon(B)+\Delta \epsilon(AB) \mbox{ },
\end{eqnarray*}
the non-additive part $\Delta \epsilon(AB)$ being the two-body increment.
For three groups of orbitals one obtains
\begin{eqnarray*}
\epsilon(ABC)=\epsilon(A)+\epsilon(B)+\epsilon(C)+\\ \Delta \epsilon(AB)+
\Delta \epsilon(BC)+\Delta \epsilon(CA)+\Delta \epsilon(ABC).
\end{eqnarray*}
Obviously, this series can be extended to any order of increments and for
the correlation energy of the solid one obtains
\begin{equation}
\label{epsilonbulk}
\epsilon_{\rm bulk} = \sum_A \epsilon(A) + \frac{1}{2} \sum_{A,B} \Delta\epsilon(AB) + \frac{1}{3!} \sum_{A,B,C}
\Delta\epsilon(ABC) + ...
\end{equation}
which is still exact. The approximation consists in neglecting
 higher than third orders 
of increments 
which makes it necessary that the sum of these increments should be 
negligibly small. Moreover, it is assumed that 
the increments rapidly decrease with increasing 
distance between groups A and B. Since 
the calculations are performed on embedded clusters, 
the
increments should only weakly depend on the chosen cluster.
The validity of these assumptions has to be checked, of course, for each
specific case. A prerequisite is that
the correlation method must be size-extensive. 
Since the method must be applicable to all types of wavefunctions
(including low-spin), we chose the quasidegenerate
variational perturbation theory (QDVPT) \cite{QDVPT}.
In the simplest case of a single-reference wavefunction, 
the correlation energy is obtained from the functional 
\begin{eqnarray*}
F[\Psi_c]=\frac{< \Psi_{SCF}+\Psi_c|H-E_{SCF} |\Psi_{SCF}+\Psi_c >}
{< \Psi_{SCF}|\Psi_{SCF}>}
\end{eqnarray*}
where $\Psi_{SCF}$ is the self-consistent field wavefunction 
and $\Psi_c$ the  
correlation part obtained by single and double substitutions from
the SCF determinant, i.e.,
\begin{eqnarray*}
|\Psi_c>= \sum_{a \atop r}c_{a}^{r}a^+_r  a_a|\Psi_{SCF}> +
\sum_{a<b \atop r<s}c_{ab}^{rs}a^+_r a^+_s a_a a_b|\Psi_{SCF}>. \mbox{ }
\end{eqnarray*}
The method is closely related to the coupled electron pair approximation
(CEPA(0))\cite{CEPA} and the averaged coupled-pair functional (ACPF)
\cite{ACPF}, see the discussion in Ref. \onlinecite{QDVPTCEPAACPF}. 
The largest increments  were compared with those obtained
from the coupled-cluster approach
with single and double substitutions (CCSD)\cite{CCSD} where
an exponential ansatz for the correlated wavefunction is made:
\begin{eqnarray*}
|\Psi_{CCSD}>=\mbox{exp}({\sum_{a \atop r}c_a^r a^+_r a_a +
\sum_{a<b \atop r<s}c_{ab}^{rs}a^+_r a^+_s a_a a_b})|\Psi_{SCF}>.
\end{eqnarray*}
These increments were also calculated
with CCSD(T)\cite{Raghavachari} where three particle
excitations are included
by means of perturbation theory.
All the calculations were done with the {\em ab-initio} program package 
MOLPRO
\cite{KnowlesWerner,MOLPROpapers,QDVPTCEPAACPF}.

\subsection{Pseudopotentials and basis sets} 
The calculations are performed on clusters of two or three ions which are 
accurately treated by quantum-chemical methods, embedded in 
surroundings treated at a lower level of approximation.
To model the Pauli repulsion of the surroundings on the clusters
considered, a pseudopotential for Ni$^{2+}$ was employed
simulating a fixed [Ar] 3$d^8$ configuration, cf. Table
\ref{PPtabelle}. The parameters have been
adjusted to averaged valence energies of 12 configurations of Ni and Ni$^+$. 
The valence energies were taken
from relativistic HF calculations using the Wood-Boring
Hamiltonian \cite{WoodBoring}.
The clusters were modeled in the same way as described in previous work on 
MgO and CaO \cite{DDFS}. Cluster O$^{2-}$ ions were surrounded by 
Ni$^{2+}$ pseudopotentials as next neighbors. 
The whole system was
embedded in a set of point charges (typically 7 $\times$ 7 $\times$ 7 
lattice sites with charges $\pm 2$ in the interior and reduced by factors
of 2, 4 and 8 at the surface planes, edges and corners, respectively). 
The description of the ions contained within the clusters is as follows.
The oxygen 
basis set was a  $[5s4p3d2f]$-basis 
\cite{Dunning}. For the nickel ions 
a pseudopotential with a Ne core \cite{DolgPP} together with 
an optimized basis set described in Ref. \onlinecite{DolgPP} 
was used. The $d$-contraction was
slightly relaxed (the three smallest exponents were used uncontracted) and
three $f$-exponents (0.67,2.4,7.8) were added yielding a $[6s5p4d3f]$ basis
set.
The correlated orbitals are $2s$ and $2p$ for O, O$^{2-}$  and the 
$3d$-orbitals in the case of Ni$^{2+}$. The doubly occupied orbitals
were kept closed in the 
reference wavefunction which leads to a single reference calculation except for
the case of low-spin coupling.
The cohesive energy
was calculated with respect to the Ni $3d^{8}4s^{2}$ state (of course,
the 4$s$ orbital was correlated in that case, too).
The influence of the correlation of the $3s$ and $3p$
orbitals of Ni will be discussed separately. 

\subsection{Test calculations}

\subsubsection{Nickel ionization potential}
Results of test calculations for the
first and second ionization potential  of Ni
are shown in Table \ref{ip}.
We performed
calculations with a finite basis set (indicated as 'our basis set') as well
as, when possible,
 using numerical schemes which are comparable to a complete basis set
(indicated as 'finite differences').
Comparing Hartree-Fock and Dirac-Fock calculations, one finds that 
relativistic effects make a significant difference. The origin and
importance of these effects on the results for the 
crystal binding energy will be
discussed later. The HF calculation with the Ne-core pseudopotential
and the $[6s5p4d3f]$ basis set is in nice agreement with the
Dirac-Fock result which shows that the pseudopotential is well suited
to account for relativistic corrections. At the correlated level,
we find that we recover most of the missing energy. Correlation 
of Ni $3s$ and $3p$ orbitals slightly improves the result; the remaining
error is mostly due to  basis-set deficiencies. The density functional
results tend towards an overestimation. 

\subsubsection{Electron affinity of oxygen}
The next test concerns the electron affinity of the oxygen atom.
We find a value of 1.359 eV at the  QDVPT
 level (exp.: 1.461 eV 
\cite{CRC}). Note that with several
density functionals negatively charged ions
are not stable because of the wrong asymptotic behavior of the functionals
(see the discussions in Ref. \onlinecite{Shore,Schwarz,GraboGross}).
Indeed, we found positive orbital energies when applying LDA and BLYP
to the O$^-$ ion which indicates the instability.

\subsubsection{Ground state of the NiO molecule}
A final test calculation was performed for the $^3\Sigma^-$ 
ground state of the
NiO molecule (see Table
\ref{niomolekel}). Our 
multi-reference averaged coupled-pair functional (MR-ACPF)
result for D$_e$ is in good agreement with the experimental
result. The bond length is slightly too short and therefore the 
vibrational frequency is too high (calculating the vibrational frequency at
the experimental bond length gives a value of 897 cm$^{-1}$ which is in 
better agreement with experiment). Enlarging the basis sets 
by using $g$ functions leads to a further
slight reduction of the
bond length. 
Correlating the Ni 3$p$ orbitals also reduces the bond length (by
$\sim$ 0.005 \AA). 
Therefore, an enlargement of the active space 
(the group of orbitals from which the reference determinants of the MR-ACPF
calculation are formed)
seems to be necessary.
The lowest-lying virtual orbitals not yet included in the active space
are of $\pi_x$ and $\pi_y$ type. 
However, 
inclusion of these orbitals is not feasible 
currently due to the high computational effort. Bauschlicher
\cite{Bauschlicher} added another virtual $\sigma$ orbital to the active
space, but on the other hand the $\delta$ orbitals and the O $2s$-orbital
were kept closed (i.e. correlated, but doubly occupied in the 
reference wavefunction).
This gave excellent agreement with experiment. However,
it is obvious from his work that the results are still sensitive to 
the correlation level applied
(see the comparison in Ref. \onlinecite{Bauschlicher}). 
Actually, we found that this problem is more critical in the case
of the molecule than in the calculations performed for the solid
(QDVPT calculations, e.g., are difficult to converge 
for the molecule). Note that, in contrast to 
the solid, the  NiO molecule is 
covalently bonded with charges $\pm$ 0.7 according to a Mulliken population
analysis.
The gradient-corrected DFT results  for NiO are excellent for the vibrational
 frequency and the bond length, but
for the dissociation energy one finds a significant overbinding.
This might change with improved functionals such as B3LYP or B3PW91 which are
not yet implemented in MOLPRO.

\section{Results for the solid}
\subsection{Incremental expansion and magnetic properties}
Regarding the one-body correlation energy increments (see Table
\ref{ink417}), one finds nearly the same value for the
increment O $\rightarrow$ O$^{2-}$ as in the systems MgO and CaO 
whereas the increment Ni $\rightarrow$ Ni$^{2+}$
is significantly larger than the corresponding value 
in the case of the alkaline earth oxides
\cite{DDFS}. 
As compared to Mg, the low lying $d$-orbitals led to an enhanced 
core-valence correlation and a larger increment for 
Ca $\rightarrow$ Ca$^{2+}$. 
The near-degeneracy effects in the 
Ni atom 
lead to an even higher valence correlation energy. 
The two-body increments between two oxygen ions are nearly 50 \% larger 
than those of MgO although the lattice constants are very similar.
This may be connected with an increased polarizability of O$^{2-}$ in NiO
which in turn might be related to the availability of low-lying open-shell
$d$-orbitals in Ni$^{2+}$.

As in the systems MgO and CaO, the Mulliken population analysis gives an
X $\rightarrow$ O  
charge transfer of slightly less than 2.
The Ni-O two-body increments turn out to make the most significant
correlation contribution. This is not surprising since the polarizability
of  the open-shell 
Ni$^{2+}$ ion is roughly six times higher than that of Mg$^{2+}$.
We used a special procedure to evaluate the Ni-O increments for technical
reasons. 
{\em Single external} correlation contributions to $\epsilon ({\rm O}
^{2-})$
involving single substitutions from localized O$^{2-}$ $2s$, $2p$ orbitals
into unoccupied orbitals, were not directly calculated for the embedded
Ni$^{2+}$ O$^{2-}$ cluster but were transferred from a calculation for a
single O$^{2-}$ embedded in large-core Ni$^{2+}$ pseudopotentials.

Analyzing the 
correlated wavefunction in the case of 
the Ni-O increments for next neighbors, 
one finds that {\em internal} excitations (excitations
from the oxygen ion to the singly occupied nickel orbitals) are negligibly
small and contribute with a few $\mu$H (Hartree units, 1 H=27.2114 eV)
only. This is certainly a consequence
of the large Coulomb repulsion $U$ in the case of doubly occupied Ni 
$3d$-orbitals. Another interesting feature is the 
importance of spin-flips on the Ni site. By restricting the excitations
for the supersystem, the Ni$^{2+}$ and O$^{2-}$ ions to {\em double externals}
(only two-particle excitations to completely
unoccupied orbitals are allowed, excitations to 
orbitals already occupied with an electron of opposite spin are forbidden),
one obtains an increment of only 0.008450 $\times 6$ H 
instead of 0.011534 
$\times$ 6 H when single external excitations are additionally taken
 into account. 
Thus, the importance of 
single externals becomes obvious. The processes among the 
single externals which are the most important ones
for the increment are those where an
electron from the O$^{2-}$ ion is  excited to a singly occupied Ni orbital
and the Ni electron is excited to a virtual orbital which results in an
effective spin-flip (Fig. \ref{Konfigurationen}). 
Of course these processes are only possible
in open shell systems. In the case of the oxygen-oxygen increment for
next neighbors, neglect of single externals only slightly reduces
the increment from 0.003916 $\times$ 6 H to 0.003549 $\times$ 6 H.
This confirms that single externals have their main impact on increments
involving open shells.

The Ni$^{2+}$-Ni$^{2+}$ increments were calculated for the case of high 
($S$=2) and low spin ($S$=0) coupling (neglecting single externals).
However, the very small energy splitting between the states is sensitive 
to the chosen
cluster and increases when point charges as next neighbors are replaced by
explicitly treated oxygens. Therefore, the results in Table \ref{ink417}
should not be used for a quantitative evaluation of the splitting.
In the case of a cluster of the type
(O(0,0,0)-Ni(0,0,1)-Ni(0,0,-1)), the splitting is 138 $\mu$H at 
the HF
level and 276 $\mu$H at the correlated level including single external
excitations (the $S$=0 state is lower). 
For next neighbors, one finds a tendency
towards ferromagnetic coupling: in the cluster (O(0,0,0)-Ni(0,0,1)-Ni(0,1,0)),
the $S$=2 state is 19 $\mu$H lower at the HF level and 22 $\mu$H when
correlations are included. These results are in qualitative agreement
with experiment where the exchange coupling is ferromagnetic 
for next neighbors
and antiferromagnetic for second next neighbors \cite{LandoltMagnetismus}. 
This indicates that the 
stability of the antiferromagnetic AF$_2$ state
(a state with parallel spins in (1 1 1) planes, adjacent planes having
antiparallel spin) found at the HF level, will be enhanced due to correlations.
Although quantitative values for the exchange coupling
 could be calculated for molecules and are in agreement with 
experiment (see, e.g., Ref. \onlinecite{Staemmler}),
the situation is more difficult in solids and the exchange coupling increases
with the number of ions explicitly treated \cite{Illas,deGraaf}. 
Therefore, our results can only give a  qualitative  picture here.
Note that the
order of magnitude of the magnetic splitting 
(0.03 eV/ double cell at the HF level) is much smaller than the
total cohesive energy.

The incremental expansion for the correlation contribution to the
cohesive energy
is well convergent. The two-body increments
show a rapid van der Waals like decrease and the sum of the three-body
corrections is small compared to the sum of one- and two-body increments.
It is interesting to see that the two-body increments amount to roughly 
80\% of the correlation contribution to the cohesive energy, while
the intra-ionic
contributions contained in the one-body increments yield only $\sim$
20\% (in
the case of the alkaline earth systems, both contributions were roughly
 of the same size).
The QDVPT results are
closer to CCSD(T) than to  CCSD, but the results are weakly
dependent only on the method applied.
Individual increments and their total contribution to the bulk cohesive
energy for a lattice spacing of 4.17 \AA \mbox{ }
are displayed in Table \ref{ink417} and Figure \ref{Niosummebild}.

\subsection{Comparison with experiment}
Our reference value for the HF binding energy (6.2 eV) is taken  from
Ref. \onlinecite{Towler}. 
However, two corrections are necessary: since the calculations of Ref. 
\onlinecite{Towler} were done
 within the unrestricted Hartree-Fock (UHF) approximation
for the solid, but at the restricted Hartree-Fock (RHF) level for the atoms 
\cite{Harrisonprivcomm}, a part of the correlation energy is already included
in that value.
 We corrected this by subtracting the energy
difference $|E_{UHF}-E_{RHF}|$  for the Ni$^{2+}$ ion (-0.0072 H) 
from the binding energy of Ref. \onlinecite{Towler}. This 
correction can be viewed as a 'one-body correction'
which is tantamount to applying Eq. \ref{epsilonbulk}
not to the correlation energy but to the difference UHF-RHF. Since the 
correction is nearly the same for the case of a system with a Ni$^{2+}$ and
an O$^{2-}$ ion, we conclude that the 'two-body corrections' are negligible
and the HF cohesive energy of Ref. \onlinecite{Towler}
 should be reduced  by 0.007 H. A second 
correction is necessary, since the CRYSTAL calculation was done neglecting
effects arising from relativity. These effects gain importance
beginning with the first row of transition metals
\cite{Relativistik}. We found (see Table \ref{ip})
 that the first and second ionization 
potentials of Ni increase
from 0.840 H (Hartree-Fock calculation) to 0.853 H 
(Dirac-Fock calculation). This is a consequence
of the relativistic contraction (and  stabilization) of the Ni
4$s$-orbital. 
In the case of Ca, this correction is of the order of 2.4 mH only.
Spin-orbit effects are negligible for Ni as the $d$-occupation in the
solid is $d^8$ ($t_{2g}^6 e_{g}^2$) and very similar to that of 
the isolated atom 
($^3F$).
Taking both corrections into account, we obtain a cohesive energy of
5.65 eV (5.59 eV at 4.17 \AA, using the bulk modulus from Ref. 
\onlinecite{Towler}). 
Adding the zero-point vibrations (0.12 eV, 
a Debye approximation with $\Theta \approx 640K$ \cite{Debye} was used)
to the experimentally determined cohesive energy of 9.5 eV \cite{CRC}, 
we thus deduce an
'experimental' correlation contribution to the cohesive energy of 
$\approx$ 4.0 eV. Within the basis sets used, we recover 3.38 eV ($\hat =$
84 \%) of this energy with our local correlation scheme. 

\subsection{Lattice constant}
By performing the incremental expansion for a second value of the
lattice constant $a$, 
we obtain the correlation correction to $a$. The same
physical picture as in the case of MgO and CaO is found: the van der Waals
like interaction between the ions (included in the two-body increments)
reduces the lattice constant whereas correlations in the oxygen ion 
tend to enhance the
lattice constant. The latter effect is
 due to the lower level spacing and therefore increasing
importance of correlations at larger lattice constant. 
The results 
for $a$=4.264 \AA \mbox{ } are displayed in Table \ref{ink4264}. 
Concerning the three-body increments, we used the two largest ones to
 evaluate the lattice constant.
By linearly fitting the correlation
energy and using the bulk modulus from the CRYSTAL calculation \cite{Towler}
(214 GPa),  we obtain a value of 
4.167 \AA, i.e., a reduction by 0.097 \AA \mbox{ } with respect to
 the CRYSTAL-HF result.
This is in nice agreement with the experimental result of 4.17 \AA 
\cite{Landolt}. 
We assumed a NaCl structure and did not take into
account the slight rhombohedral distortion observed in the experiment.
The calculated cohesive energy at this equilibrium constant is
(5.65-0.06+3.38) eV=8.97 eV which is 93 \% of the total cohesive energy of
9.6 eV.

Finally, we evaluated the one-body and two-body Ni-O increments when 
electrons in 
the Ni $3s$ and $3p$ orbitals are correlated. At a lattice constant
of 4.17 \AA, we find that the one-body increment Ni $\rightarrow$ Ni$^{2+}$
increases from 0.074545 to 
0.080790 as a consequence of the outer-core- valence correlation.
On the other hand, the absolute value of the 
sum of the Ni-O increments increases from 0.073446 to 0.082706. 
As these effects have the opposite sign, they
nearly cancel and increase the correlation contribution to the cohesive 
energy to 0.127204 H at 4.17 \AA. The lattice constant is reduced 
to 4.151 \AA. 

\subsection{Comparison with results from literature}
Our results for the cohesive energy are in better agreement with 
experiment than those
of DFT, where a significant overestimation was found 
\cite{Yamashita}.
 The lattice constant of the latter work (4.183 \AA) is in good
agreement with experiment.
A recently performed Quantum Monte-Carlo (QMC) calculation \cite{Tanaka}
gives good agreement with  experiment, but 
the error-bars are quite large compared to those of 
other QMC work \cite{QMC}.
 Moreover, the basis sets
are very small and not adjusted to the pseudopotential used.

\section{Conclusion}
In conclusion, we have shown that the incremental scheme is capable of
accurately
treating systems with open $d$-shells. 
We obtained more than 84 \% of the correlation contribution to the cohesive
energy or 93 \% of the total  cohesive energy. 
Contributions from interatomic origin are roughly four times larger than
those due to the formation of ions.
The lattice constant agrees
well with experiment. The stability of the magnetic 
ground state obtained found at the HF level 
is enhanced with the inclusion of correlations.
We feel that the quality
of this calculation is superior to previous work with DFT and QMC.

\acknowledgments
We are grateful to Prof.\ H.-J.\ Werner (Stuttgart) for providing the
program package MOLPRO and to Dr. N. M. Harrison (Daresbury) for useful
information on the CRYSTAL calculations.

\newpage
\begin{figure}
\caption{Spin ordering: a) 
in the ground state of NiO; b) for a double external 
configuration with simultaneous excitations from a Ni$^{2+}$ and an 
O$^{2-}$ ion; c) for a 
single external configuration with spin-flip.}
\vspace{0.5cm}
\label{Konfigurationen}
\end{figure}
\mbox{ }

\begin{figure}
\caption{Result of the incremental expansion for NiO at a lattice constant
of 4.17 \AA.}
\vspace{0.5cm}
\label{Niosummebild}
\end{figure}

\newpage
\begin{table}
\begin{center}
\caption{\label{PPtabelle}
Parameters of the Ni$^{2+}$ pseudopotential. The pseudopotential
has the form $V(r)=-\frac{Q}{r}+\sum_{l}\sum_{k}A_{kl}exp
(-a_{kl} r^2)\sum_{m_l}|l\mbox{{$m_l$}$><$}lm_l|.$}
\vspace{5mm}
\begin{tabular}{|ccccc|}
Q & $l$ & $k$ & $A_{kl}$ & $a_{kl}$\\ \hline
2 & 0 & 1 & 6.3098777 & 0.8926653 \\
& 0 & 2 & -0.7025295 & 0.3059013 \\
& 1 & 1 & 3.2304525 & 0.7566913 \\
& 1 & 2 & -0.2278919 & 0.2415367 \\
& 2 & 1 & 1.1107680 & 0.3984276 \\
& 3 & 1 & -0.5915083 & 0.4892060 \\
\end{tabular}
\end{center}
\end{table}

\begin{table}
\begin{center}
\caption{\label{ip}Sum of first and second ionization potentials for Ni
in Hartree units ($^3$F state)}
\vspace{5mm}
\begin{tabular}{|c|c|}
 &  Ni $\rightarrow$ Ni$^{2+}$ \\
\hline
Hartree-Fock, finite differences & 0.840310  \\
Dirac-Fock, finite differences & 0.852950 \\
Hartree-Fock with relativistic pseudopotential, & \\
our basis & 0.855544 \\
QDVPT (3$d$, 4$s$ correlated), our basis & 0.930863 \\
QDVPT (3$s$, 3$p$, 3$d$, 4$s$ correlated), our basis & 0.936216 \\
LDA, our basis & 0.999031 \\
BLYP, our basis & 0.982560 \\
expt. & 0.947760 \cite{Moore} \\
\end{tabular}
\end{center}
\end{table}

\begin{table}
\begin{center}
\caption{\label{niomolekel}
Bond length $R_e$ (\AA), 
dissociation energy $D_e$ (eV) and vibrational frequency
$\omega_e$ (cm$^{-1}$) of the NiO
molecule. The orbitals correlated with the multi-reference 
averaged coupled-pair functional (MR-ACPF) 
scheme
are O $2s$, $2p$, Ni $3d$, $4s$. The 
dissociation energy is calculated with respect to the $^3D$ $(d^9s^1)$ state 
of Ni. The DFT calculations for the atoms were performed without applying a
spherical approximation for the potential and with integer occupation numbers.}
\vspace{5mm}
\begin{tabular}{|c|ccc|}
 &  $R_e$ & \multicolumn{1}{c}{$D_e$} & $\omega_e$ \\ \hline
MR-ACPF & 1.601 & 3.62 & 958 \\
LDA & 1.584 & 6.12 & 932\\
BLYP & 1.626 & 4.73 & 861 \\
Ref. \onlinecite{Bauschlicher} 
& 1.626 &  3.80 & 850  \\
expt. & 1.627 \cite{Srdanov} & 3.92$\pm$0.03 \cite{Watson} & 838 
\cite{Srdanov,Green}\\
\end{tabular}
\end{center}
\end{table}

\begin{table}
\begin{center}
\caption{\label{ink417}Local correlation energy 
increments (in H) at a lattice constant of 4.17 \AA, from QDVPT calculations.
The largest increments are compared with CCSD and CCSD(T).
 The numbers in brackets (a,b,c) are
indices of the lattice site of the ion and are
given in multiples of the lattice constant. $S$ denotes the spin coupling
($S=2$ means ferromagnetic, $S=0$ antiferromagnetic coupling). The 
increments are already multiplied with the weight factor except where the
weight factor is explicitly repeated in the sum. A weight factor of 0 means
that this state does not occur in the  AF$_2$ state. In the case
of the three-body increments, we summed only over the two largest ones
in order to make a
comparison with the results of Table 
V 
possible.}

\vspace{5mm}
\begin{tabular}{|c|c|c|}
 & weight 
 & \multicolumn{1}{c|}{QDVPT} 
\\
\hline
$\rm {Ni} \rightarrow \rm {Ni^{2+}}$ & 1 & +0.074545 
 \\
CCSD: +0.068628 & & \\
CCSD(T) : +0.073366 & &  \\

$\rm {O}  \rightarrow \rm {O^{2-}}$ & 1 &   -0.100823 \\
CCSD : -0.099578 & &  \\
CCSD(T): -0.105364 & & \\
\hline
\multicolumn{2}{|c|} {sum of one-body increments}   & -0.026278 \\
\hline
\rm{Ni(0,0,0)-O(1,0,0)} 
& 6 & -0.069204
\\

CCSD: -0.059748 & & \\
CCSD(T): not possible & & \\

\rm{Ni(0,0,0)-O(1,1,1)} & 8 & -0.002616 
\\
\rm{Ni(0,0,0)-O(2,1,0)} & 24 & -0.001392
\\
\rm{Ni(0,0,0)-O(3,0,0)} & 6 & -0.000042
\\
\rm{Ni(0,0,0)-O(2,2,1)} & 24 & -0.000192  \\
\hline
\multicolumn{2}{|c|} {sum of Ni-O increments}   
 & -0.073446 \\
\hline
\rm{Ni(0,0,1)-Ni(0,1,0)}, $S$=2 & 3 & -0.000831
\\
\rm{Ni(0,0,1)-Ni(0,1,0)}, $S$=0 & 3 & -0.000780
\\
\rm{Ni(1,0,0)-Ni(-1,0,0)},  $S$=2 & 0 & -0.000025 $\times$ 0 
\\
\rm{Ni(1,0,0)-Ni(-1,0,0)}, $S$=0 & 3  & -0.000025 $\times$ 3 
\\
\hline
\multicolumn{2}{|c|} {sum of Ni-Ni increments}   
 & -0.001686 \\
\hline
\rm{O(0,0,0)-O(0,1,1)} & 6 & -0.023496 
\\

CCSD: -0.020136 & & \\
CCSD(T): -0.023928 & & \\

\rm{O(0,0,0)-O(2,0,0)} & 3 & -0.001155 
\\
\rm{O(0,0,0)-O(2,1,1)} & 12 & -0.001200 
\\
\rm{O(0,0,0)-O(2,2,0)} & 6 & -0.000228 
\\
\rm{O(0,0,0)-O(3,1,0)} & 12 & -0.000228
\\
\rm{O(0,0,0)-O(2,2,2)} & 4 & -0.000044 \\
\hline
\multicolumn{2}{|c|} {sum of O-O increments}   
 & -0.026351 \\
\hline
\rm {O(1,0,0)-O(0,1,0)-O(0,0,1)} 
 & 8 & +0.000848
\\
\rm {O(1,0,0)-O(-1,0,0)-O(0,0,1)}
 & 12 & +0.000252
\\
\rm {O(0,0,0)-Ni(1,0,0)-Ni(0,1,0),$S$=2} 
 & 6 & +0.000234 
\\
\rm {O(0,0,0)-Ni(1,0,0)-Ni(0,1,0),$S$=0} 
 & 6 & +0.000222 
\\
\rm {O(0,0,0)-Ni(0,0,1)-Ni(0,0,-1),$S$=2}
 & 0 &  -0.000028  $\times$ 0
\\
\rm {O(0,0,0)-Ni(0,0,1)-Ni(0,0,-1),$S$=0}
 & 3 & -0.000049 $\times$ 3 
\\
\rm {O(0,0,0)-Ni(0,1,0)-O(0,1,1)}
  & 12 & +0.002724 
\\
\rm {O(1,0,0)-O(-1,0,0)-Ni(0,0,0)}
 & 3 & +0.000078 
\\
\hline
\multicolumn{2}{|c|} {sum of two largest three-body increments} & +0.003572
\\
\hline
\multicolumn{2}{|c|} {total sum} & -0.124189
\\
\hline
\multicolumn{2}{|c|} {expt.} & -0.147 
\\
\end{tabular}
\end{center}
\end{table}

\begin{table}
\begin{center}
\caption{\label{ink4264}
Local increments (in H) for NiO at a lattice constant of 4.264 \AA.}
\vspace{5mm}
\begin{tabular}{|c|c|}
 & \multicolumn{1}{c|}{QDVPT} 
\\
\hline
\multicolumn{1}{|c|} {sum of one-body increments}   
& -0.028014
\\
sum of Ni-O increments& -0.068362 \\
sum of Ni-Ni increments & -0.001416 \\
sum of O-O increments & -0.025207 \\
\multicolumn{1}{|c|} {sum of three-body increments} & +0.003088 \\
\hline
\multicolumn{1}{|c|} {total sum} & -0.119911
\\
\end{tabular}
\end{center}
\end{table}

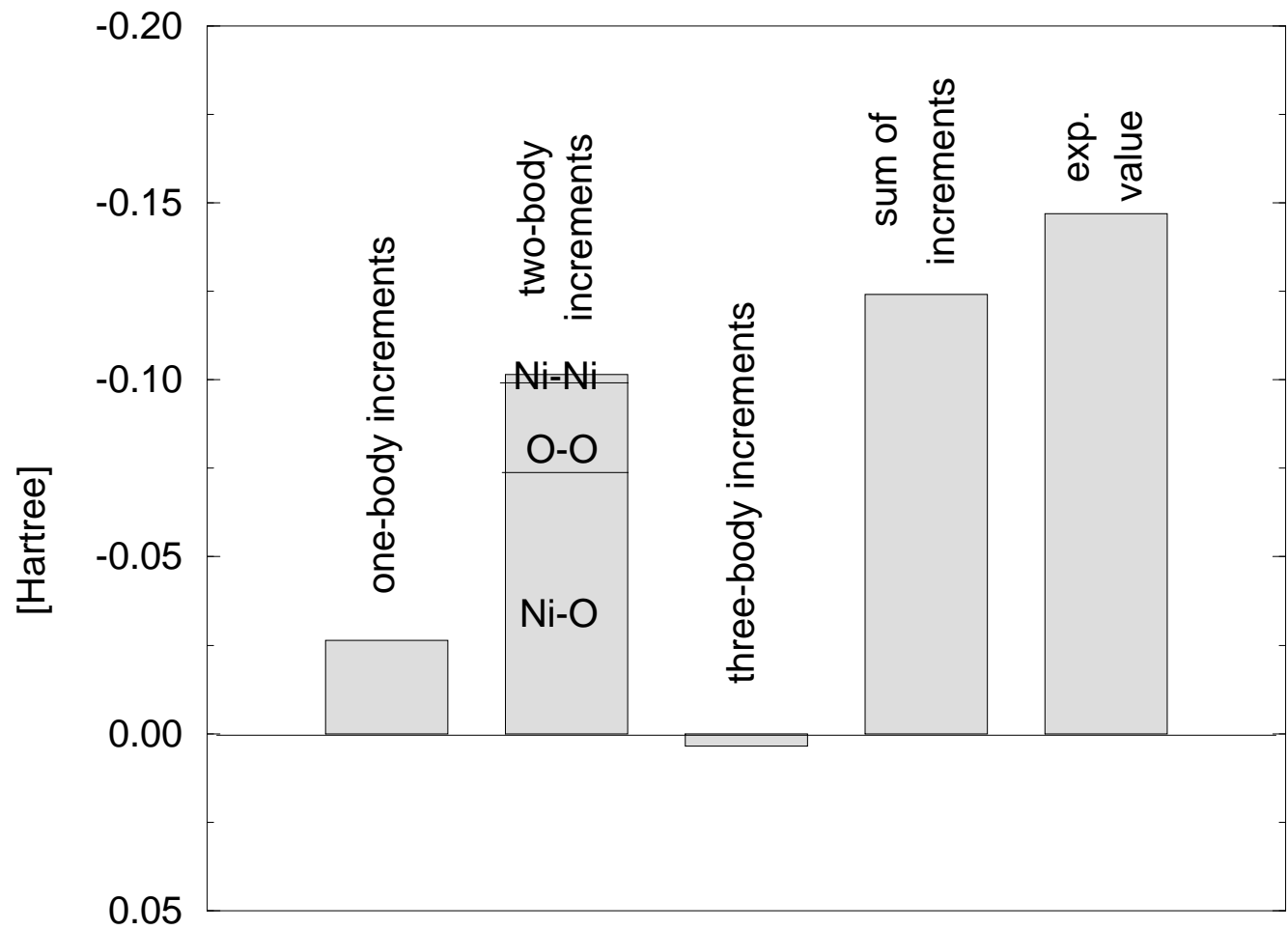
\end{document}